\documentclass[apj,numberedappendix]{emulateapj}
\usepackage{apjfonts}

\begin{document}

\title{Spreading of Accreted Material on White Dwarfs}

\author{Anthony L. Piro}
\affil{Department of Physics, Broida Hall, University of California
	\\ Santa Barbara, CA 93106; piro@physics.ucsb.edu}

\and

\author{Lars Bildsten}
\affil{Kavli Institute for Theoretical Physics and Department of Physics,
Kohn Hall, University of California
	\\ Santa Barbara, CA 93106; bildsten@kitp.ucsb.edu}

\begin{abstract}

  When a white dwarf (WD) is weakly magnetized and its accretion disk is thin,
accreted material first reaches the WD's surface at its equator. This matter
slows its orbit as it comes into co-rotation with the WD, dissipating kinetic
energy into thermal energy and creating a hot band of freshly accreted material
around the equator. Radiating in the extreme ultraviolet and soft X-rays, this
material moves toward the pole as new material piles behind it, eventually
becoming part of the WD once it has a temperature and rotational velocity
comparable with the surface. We present a set of solutions which describe
the properties of this ``spreading layer'' in the steady state limit based
on the conservation equations derived by Inogamov \& Sunyaev (1999) for
accreting neutron stars. Our analysis and subsequent solutions show that the
case of WDs is qualitatively different. We investigate example solutions of
the spreading layer for a WD of mass $M=0.6M_\odot$ and radius
$R=9\times10^{8}\textrm{ cm}$. These solutions show that the spreading layer
typically extends to an angle of $\theta_{\rm SL}\approx0.01-0.1$
(with respect to the equator), depending on accretion rate and the
magnitude of the viscosity. At low accretion rates,
$\dot{M}\lesssim10^{18}\textrm{ g s}^{-1}$, the amount of spreading is
negligible and most of the dissipated energy is radiated back into the
accretion disk. When the accretion rate is high, such as in dwarf novae,
the material may spread to latitudes high enough to be visible above the
accretion disk. The effective temperature of the spreading layer is
$\sim(2-5)\times10^5\textrm{ K}$ with approximately
$T_{\rm eff}\propto\dot{M}^{9/80}$. This power-law dependence on $\dot{M}$
is weaker than for a fixed radiating area and may help explain extreme
ultraviolet observations during dwarf novae. We speculate about other high
accretion rate systems ($\dot{M}\gtrsim10^{18}\textrm{ g s}^{-1}$) which
may show evidence for a spreading layer, including symbiotic binaries and
supersoft sources.
\end{abstract}


\section{Introduction}

  In close binaries containing a white dwarf (WD), the companion star can
fill its Roche lobe either through stellar evolution or angular momentum
loss and begin accreting onto the WD to form a cataclysmic variable
(CV; Warner 1995). Before material can reach the WD it must transport angular
momentum outward via an accretion disk. Observations and theoretical studies
show that when the disk can radiate the internally dissipated energy
(Pringle 1981) it will be thin (i.e. the vertical height of the disk is
much smaller than the WD radius; Shakura \& Sunyaev 1973;
Meyer \& Meyer-Hofmeister 1982) and rotating at nearly Keplerian velocity.
If the WD is weakly magnetized, the accreted material reaches the surface at
the equator and then must pass through a transition region to settle into
the WD and become part of the star. When the WD spin is much less breakup,
nearly half of the accretion
luminosity is released in this region, making it as luminous as the accretion
disk and likely crucial to understanding the luminosity from accreting WDs.

  The transition from the accretion disk to the WD surface is typically
treated in a boundary layer (BL) model, which is simply an extension
of the accretion disk with additional torques provided by the WD to
decelerate disk material to the WD rotational velocity. In BL studies
the important coordinate is assumed to be radial and the vertical
structure is solved by assuming a vertical pressure scale height
(e.g. Popham \& Narayan 1995). The geometry of the BL is of a fattened
disk near the WD surface. Other BL studies solve the two-dimensional problem
simultaneously in both radial and latitudinal directions using numerical
techniques (e.g. Kley \& Hensler 1987; Kley 1989a, 1989b).
Kippenhanh \& Thomas (1978) investigated the properties of ``accretion
belts'' on WDs by assuming that the accreted material is
marginally stable to the Richardson criterion. We now want to
understand the state of recently accreted material on much shorter
timescales ($\ll10^{4}\textrm{ yr}$), and assess the effects of a turbulent
viscous stress on the spreading properties.
In Inogamov \& Sunyaev (1999; hereafter IS99), this problem was approached from
a new angle to study accreting neutron stars (NSs). Their method follows
the latitudinal flow of matter on the compact object which
provides information about the spreading area of hot, freshly accreted
material which is not captured in BL models. We now apply these same methods
to the case of WDs and call this model the spreading layer (SL), to
differentiate it from BL studies. We solve the conservation equations derived
by IS99 for fluid confined to the WD surface and find that the steady state
limit naturally attracts toward solutions where local viscous dissipation
is nearly balanced by local radiative cooling. This is qualitatively
different than the results of IS99 who found that the SLs of NSs show a
great deal of advection of dissipated energy up to higher latitudes where
it is radiated. This is mainly due to their high accretion rates near the
Eddington limit.

  For a fiducial WD of mass $M=0.6M_\odot$ and radius
$R=9\times10^{8}\textrm{ cm}$, our solutions yield a thin hot band of
extent $\theta_{\rm SL}\approx0.01-0.1$ with
an effective temperature of $\sim(2-5)\times10^5\textrm{ K}$ over a range
of accretion rates from $10^{17}-10^{19}\textrm{ g s}^{-1}$, implying that
the SL should contribute to accreting WD spectra in the extreme
ultraviolet (EUV) or soft X-rays. We also find the SL's dependence on
accretion rate and viscosity. These solutions show that at low accretion
rates ($\dot{M}\lesssim10^{18}\textrm{ g s}^{-1}$) the SL does not appear
above the accretion disk, so that radiation from the SL may be obscured by
the accretion disk or its winds. Only when
$\dot{M}\gtrsim10^{18}\textrm{ g s}^{-1}$, such as in dwarf novae,
symbiotic binaries, or supersoft sources, will the SL reach high enough
latitudes to be directly seen. Perhaps the best
chance for testing the SL model are dwarf novae in outburst and superoutburst.
Some observations show a fairly constant effective temperature over a
large change in luminosity (for example SS Cyg; Mauche, Raymond \& Mattei 1995)
which is consistent with a SL. Direct comparisons between our simple model
and observations can be complicated if the SL emission is absorbed and
re-radiated by accretion disk winds, as is likely for systems
which show that the EUV is not eclipsed by the secondary whereas the WD is
(such as for the edge-on OY Car in superoutburst; Mauche \& Raymond 2000).
Nevertheless, the SL may be important for calculating the underlying continuum
scattered by these winds. We predict a quick timescale for spreading,
$t_{\rm SL}\sim50\textrm{ s}$, which means that the typical mass of the
spreading region ($\lesssim10^{20}\textrm{ g}$) is much less than the
``accretion belt'' of Kippenhahn \& Thomas (1978). Most likely, as angular
momentum is tranferred between the SL and the underlying WD surface, the
underlying surface layers will spin up similar to what Kippenhahn \&
Thomas (1978) calculate. The short spreading timescale also means that
following a dwarf nova outburst, when accretion onto the WD has decreased
dramatically, the SL will spread over the WD surface too quickly to be
observed as a separate hot component.

  In \S 2 we solve for the radial structure of the SL using a one-zone model.
We derive the differential equations which describe the SL in the steady state
limit in \S 3, using IS99 as a guide. In \S 4 we discuss the solutions
of the spreading equations. We study physically motivated analytic estimates
which capture the essential features of the numerical integrations. These
provide power law relations which show how the SL changes with accretion
rate, viscosity, and the properties of the accreting WD. We then present
the results of integrating the equations numerically to undestand the
SL in more detail. In \S 5 we discuss observational tests for the SL
and consider the current data in relation to our calculations. We conclude
in \S 6 with a summary of our findings along with a discussion about further
work which can be done with this model.
 
\section{Radial Properties of the Hot Spreading Layer}

  Following IS99, we first construct a one-zone model to capture the
radial extent of the hot, rapidly rotating material in the SL, derived
in a frame on the surface of a WD. The SL is assumed to start near the
equator in hydrostatic balance. The pressure scale height,
$h=P/\rho g \sim 10^7\textrm{ cm}$ (this will be shown in \S 4),
is significantly less than the radius of a typical WD so we assume a
plane parallel geometry with constant gravitational acceleration. Initially
the azimuthal velocity of the spreading material, $v_\phi$, will be nearly
Keplerian, $v_\phi\approx v_{\rm K}\equiv(GM/R)^{1/2}$, so that gravitational
acceleration can be significantly decreased by centrifugal effects. We
therefore use an ``effective'' gravitational constant
\begin{eqnarray}
        g_{\rm eff} = \frac{GM}{R^2}
                - \frac{v_\theta^2}{R}
		- \frac{v_\phi^2}{R},
\end{eqnarray}
where $v_\theta$ is the latitudinal velocity, $\theta$ is measured with
respect to the equator, and we assume that the WD is rotating much slower
than breakup. Since $g_{\rm eff}$ is independent of radius in the thin limit,
hydrostatic balance is integrated to give
\begin{eqnarray}
	P=g_{\rm eff}y,
\end{eqnarray}
where $y$ is the column depth, defined as $dy\equiv-\rho dz$, and $z$ is the
radial coordinate in this plane parallel layer.

  Radiation pressure may play an important role in the hot SL
so we define the force per unit mass due to radiation flux
at the photosphere,
\begin{eqnarray}
        g_{\rm rad} = \frac{\kappa F}{c},
\end{eqnarray}
where $\kappa$ is the opacity, $F$ is the flux, and $c$ is the speed of
light. If the one-zone layer becomes optically thin we cannot assume the
flux provides this force and these approximations become invalid.
With these definitions $g_{\rm eff}=g_{\rm rad}$ defines the local
Eddington accretion rate. We use an opacity dominated by Thomson scattering
($\kappa=0.34 \textrm{ g cm}^{-2}$ for a solar composition with mean
molecular weight per electron of $\mu_e\approx1.2$). Free-free absorption
may also be an important mechanism for radiative transport so we check
our assumption of an electron scattering dominated
opacity once we find the numerical solutions for the SL in \S 4.

  We assume that the flux is constant throughout the layer. Implicit
in this assumption is that the majority of the dissipation is at the
bottom of the layer where hot, rapidly rotating spreading material
comes into contact with the cold, slowly rotating WD. Radiative
diffusion results in
\begin{eqnarray}
	F = \frac{4acT^3}{3\kappa}\frac{dT}{dy},
\end{eqnarray}
where $a$ is the radiation constant.

  Pressure is related to density and temperature by the equation of state,
\begin{eqnarray}
        P = \frac{\rho kT}{\mu m_p} + \frac{aT^4}{3},
\end{eqnarray}
where $k$ is Boltzmann's constant and $\mu$ is the mean molecular weight
($\mu=0.62$ for a solar composition). Equation (4) is integrated and
combined with equation (5) to give
\begin{eqnarray}
	T &=& \left(\frac{3\kappa F}{ac}\right)^{1/4} y^{1/4},
\end{eqnarray}
and
\begin{eqnarray}
	\rho &=& \frac{\mu m_p}{kT} (g_{\rm eff}-g_{\rm rad})y.
\end{eqnarray}
We use equations (2), (6), and (7) to describe the layer, where $y$ is
the column depth at the base of the spreading matter. This material will
initially have a much higher entropy than the WD surface when it arrives at
the equator. As the material moves toward the WD pole it will radiate
and slow its rotational velocity until it becomes part of the WD (Figure 1).

\begin{figure*}
\epsscale{0.8}
\plotone{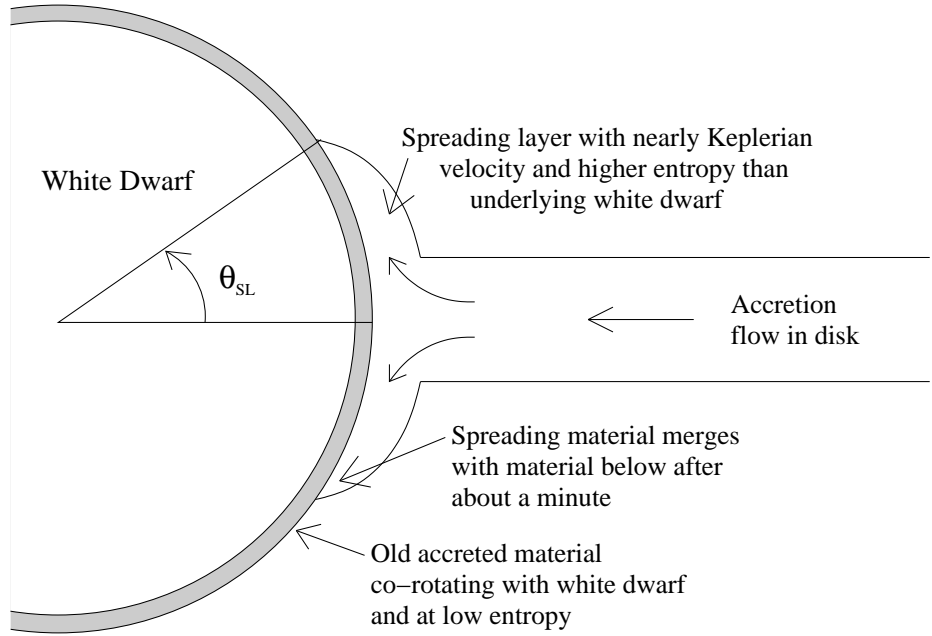} 
\figcaption{A schematic diagram of the spreading layer making contact
with the white dwarf. The accretion flow arrives at the white dwarf with
nearly Keplerian azimuthal velocity. As material piles at the equator,
rotational energy is dissipated, heating the layer. The pressure which 
is created at the equator forces material toward the pole, creating a
spreading layer. The material continues to travel away from the equator
until it loses its azimuthal velocity due to frictional dissipation and has
an entropy similar to the underlying white dwarf. At this point the
spreading material becomes part of the slowly downward advecting
material on the white dwarf surface.}
\end{figure*}

\section{Derivation of Spreading Equations}

  Since we are primarily interested in how the properties of the accreting
material change with latitude, it is easiest to consider the equations
for conserved fluxes (fluxing in the $\theta$-direction), integrated
radially through the SL. These equations are taken in the steady state
limit, resulting in a set of equations analogous to those derived by IS99.
We include the majority of the derivation here as it will assist our choice
of boundary conditions which is different than what IS99 found for NSs. In
our derivation we primarily use notation consistent with IS99, but make
changes when convenient to do so.

\subsection{Conservation Equations}

  We first derive the general, time dependent, conservation equations
in a frame on the WD. For conservation of mass we consider the radially
integrated density (i.e. the column depth, $y$),
\begin{eqnarray}
	\frac{\partial}{\partial t} (2\pi Ry\cos\theta)
	+ \frac{1}{R}\frac{\partial}{\partial\theta}
		(2\pi Ryv_\theta\cos\theta)
	= 0,
\end{eqnarray}
which simplifies to
\begin{eqnarray}
	R\cos\theta\frac{\partial y}{\partial t}
        + \frac{\partial}{\partial\theta}
                (yv_\theta\cos\theta) 
        = 0.
\end{eqnarray}
Conservation of momentum in the $\theta$-direction is given by
\begin{eqnarray}
	\lefteqn{\frac{\partial}{\partial t} (2\pi Ryv_\theta\cos\theta) 
        + \frac{1}{R}\frac{\partial}{\partial\theta} 
                (2\pi Ryv_\theta^2\cos\theta)
	+ 2\pi Ry\frac{v_\phi^2}{R}\sin\theta}\hspace{1.5cm}
	\nonumber
	\\
        &=& -2\pi R\cos\theta\frac{1}{R}\frac{\partial}{\partial\theta}
		\left(\int_0^h P dz\right)
	- 2\pi R\tau_\theta\cos\theta.\hspace{1cm}
\end{eqnarray}
which can be rewritten as,
\begin{eqnarray}
        \lefteqn{R\cos\theta\frac{\partial}{\partial t} (yv_\theta)
        + \frac{\partial}{\partial\theta}
                (yv_\theta^2\cos\theta)
        + yv_\phi^2\sin\theta}\hspace{1cm}
	\nonumber
	\\
        &=& -\cos\theta\frac{\partial}{\partial\theta}
                \left(\int_0^h P dz\right)
        - R\tau_\theta\cos\theta,
\end{eqnarray}
where the last term on the left side is a centrifugal piece which comes from
the gradient of the spherical coordinate basis vectors
(Landau \& Lifshitz 1959),
the first term on the right is the radially integrated hydrostatic force, and
$\tau_\theta$ is the viscous stress in the $\theta$-direction. The total
viscous stress, $\tau$, is from the turbulence created as high
entropy fluid quickly rotates against the lower entropy WD material below.
We parametrize $\tau$ in terms of a unitless constant, $\alpha$, so that
\begin{eqnarray}
        \tau = \alpha \rho v^2,
\end{eqnarray}
where $v^2 = v_\theta^2+v_\phi^2$.
In Appendix A we consider lower limits to $\alpha$ provided by ion or
radiative viscosities. These are found to be fairly small
($\alpha\lesssim10^{-6}$), implying that turbulence is probably the
dominant cause of friction. To be as general as possible we treat $\alpha$
as a free parameter and study its affect on the spreading properties
for the values $\alpha=10^{-4}-10^{-2}$. This range of values is chosen for
the physically realistic solutions which result for
$\dot{M}\sim10^{17}-10^{19}\textrm{ g s}^{-1}$. If $\alpha\lesssim10^{-4}$
the viscous stress is too slow to halt the spreading, and the
spreading flow diverges in numerical integrations. For $\alpha\gtrsim10^{-2}$
the dissipation becomes so large that the scale height puffs up to be
comparable to the WD radius, making the plane parallel assumptions invalid.
In such a case a more detailed study of the radial structure is required
than what we provide in \S 2. We discuss the available parameter space for
values of $\alpha$ in more detail in \S 4 when we consider the boundary
conditions. The components of $\tau$ are
\begin{eqnarray}
        \tau_\theta = \alpha\rho v^2 \frac{v_\theta}{v}
        =  \alpha\rho v_\theta\sqrt{v_\theta^2+v_\phi^2},
\end{eqnarray}
and
\begin{eqnarray}
        \tau_\phi = \alpha\rho v_\phi\sqrt{v_\theta^2+v_\phi^2},
\end{eqnarray}
in the latitudinal and azimuthal directions, respectively.

  The pressure, integrated over a scale height, $h$, is
\begin{eqnarray}
        \int_0^h P dz &=& \frac{k}{\mu m_p}
	\left( \frac{3\kappa F}{ac}\right)^{1/4}
        \frac{g_{\rm eff}}{g_{\rm eff}-g_{\rm rad}}
        \int_0^y y^{1/4} dy
        \nonumber
        \\
        &=& \frac{4}{5}
        \frac{g_{\rm eff}}{g_{\rm eff}-g_{\rm rad}} \frac{kTy}{\mu m_p},
\end{eqnarray}
where both $y$ and $T$ are assumed to be their respective values at the base
of the SL, and $\kappa$ is assumed to be constant. We see that difficulties
arise in the one-zone formulation when the SL becomes close to the Eddington
limit ($g_{\rm eff}=g_{\rm rad}$). This must be carefully considered when
we solve the resulting differential equations in \S 4.

  Next, there is conservation of momentum in the $\phi$-direction,
\begin{eqnarray}
        \frac{\partial}{\partial t} (2\pi Ryv_\phi\cos\theta)
        + \frac{1}{R}\frac{\partial}{\partial\theta}
                (2\pi Ryv_\phi v_\theta\cos\theta)\hspace{1 cm}
	\nonumber
        \\
        \hspace{2 cm}- 2\pi Ry\frac{v_\theta v_\phi}{R}\sin\theta
        = - 2\pi R\tau_\phi\cos\theta.
\end{eqnarray}
Simplifying this expression we find,
\begin{eqnarray}
        R\cos\theta\frac{\partial}{\partial t} (yv_\phi)
        + \frac{\partial}{\partial\theta}
                (yv_\phi v_\theta\cos\theta)
        - y v_\phi v_\theta\sin\theta
	\nonumber
        \\
	= - R\tau_\phi\cos\theta,
\end{eqnarray}
where $\tau_\phi$ is the viscous stress in the $\phi$-direction
as explained above.

  The last conservation equation to consider is that for energy per unit
area within the SL. This is composed of many pieces, the first being
kinetic energy,
\begin{eqnarray}
	E_{\rm KE} = \frac{1}{2}(v_\theta^2+v_\phi^2)y.
\end{eqnarray}
The gravitational potential energy is
\begin{eqnarray}
        E_{\rm grav} =  \int_0^h \rho g_{\rm eff}zdz
	= \frac{4}{5}
        \frac{g_{\rm eff}}{g_{\rm eff}-g_{\rm rad}} \frac{kTy}{\mu m_p}.
\end{eqnarray}
The internal energy of the gas is
\begin{eqnarray}
        E_{\rm gas} = \frac{3}{2}\int_0^h \frac{\rho kT}{\mu m_p} dz
	= \frac{6}{5}\frac{kTy}{\mu m_p},
\end{eqnarray}
and the radiative energy is
\begin{eqnarray}
	E_{\rm rad} = \int_0^h aT^4 dz
	= \frac{12}{5}
	\frac{g_{\rm rad}}{g_{\rm eff}-g_{\rm rad}}\frac{kTy}{\mu m_p}.
\end{eqnarray}
The total radially integrated energy density is then
$E_{\rm tot}  = E_{\rm KE}+E_{\rm grav}+E_{\rm gas}+E_{\rm rad}$. With this
we express energy conservation as
\begin{eqnarray}
	\lefteqn{\frac{\partial}{\partial t} (2\pi RE_{\rm tot}\cos\theta)
        + \frac{1}{R}\frac{\partial}{\partial\theta}
                \left[2\pi Rv_\theta\cos\theta\left(E_{\rm tot}+\int_0^h Pdz
			\right)\right]}\hspace{5cm}
	\nonumber
	\\
        &&= - 2\pi RF\cos\theta,\hspace{1cm}
\end{eqnarray}
where the extra term added to the advected energy density is the radially
integrated work that is done as fluid elements move higher in latitude.
This expression can be simplified to give
\begin{eqnarray}
        \lefteqn{R\cos\theta\frac{\partial E_{\rm tot}}{\partial t}
        + \frac{\partial}{\partial\theta}
                \left[v_\theta\cos\theta\left(E_{\rm tot}+\int_0^h Pdz
                        \right)\right]}\hspace{5cm}
	\nonumber
	\\
        &&= - RF\cos\theta.\hspace{1cm}
\end{eqnarray}
Equations (9), (11), (17), and (23) must now be solved to find the properties
of the SL.

\subsection{Taking the Steady State Limit}

  As in IS99 we solve the SL equations in the steady state limit
($\partial/\partial t=0$) which  greatly simplifies the solutions because
$\theta$ is the only independent variable. To be a valid
approximation, we require a sink for the accreted material at some
latitude above the equator. The material near the equator moves in
latitude rather quickly with $v_\theta\sim10^5-10^6\textrm{ cm s}^{-1}$
(as we show in \S 4) and thus passes through the SL on a timescale of
tens of seconds. Above this region the solutions show a long, deep
minimum in $v_\theta$ for most of the star where
$v_\theta\sim10^4\textrm{ cm s}^{-1}$. This means that fluid
still moving higher in latitude can only reach the pole
on a timescale of $\sim10^4 \textrm{ s}$. At this point the fluid
will have lost its azimuthal velocity, dissipated its high internal
energy, and have an entropy set by the underlying WD luminosity.
This means that long before the fluid has time to reach the pole it will mix
with the slowly downward advecting fluid of the WD surface and thus be part of
the WD. Recent theoretical
calculations (Townsley \& Bildsten 2003) and observations
(Sion 1999) show that accreting WDs have
effective temperatures $T_{\rm eff}$ $\approx(1-3)\times10^4\textrm{ K}$.
At a latitude at which the SL has reached a similar $T_{\rm eff}$ and
has $v_\phi\approx0$ we therefore set a sink where we stop our integrations.

  The steady state assumption implies that continuity is set by the accretion
rate, so that integrating equation (9) results in
\begin{eqnarray}
	\frac{1}{2}\dot{M} = 2\pi Ryv_\theta\cos\theta.
\end{eqnarray}
We use $yv_\theta\cos\theta  = \textrm{constant}$ and the steady state
approximation to simplify the other conservation equations. Equation (11)
becomes
\begin{eqnarray}
        yv_\theta\cos\theta\frac{dv_\theta}{d\theta}
        + \frac{4}{5}\cos\theta\frac{d}{d\theta}
		\left[
	\frac{g_{\rm eff}}{g_{\rm eff}-g_{\rm rad}}\frac{kTy}{\mu m_p}\right]
	\nonumber
	\\
        = - R\tau_\theta\cos\theta - yv_\phi^2\sin\theta.
\end{eqnarray}
whereas equation (17) is rewritten
\begin{eqnarray}
        yv_\theta\cos\theta\frac{dv_\phi}{d\theta}
	= - R\tau_\phi\cos\theta
        + yv_\theta v_\phi\sin\theta.
\end{eqnarray}
Equation (23) is simplified to
\begin{eqnarray}
        \lefteqn{yv_\theta\cos\theta\frac{d}{d\theta}
        \left[ \frac{v_\theta^2+v_\phi^2}{2}
                + \frac{2}{5}
  \frac{7g_{\rm eff}+3g_{\rm rad}}{g_{\rm eff}-g_{\rm rad}}\frac{kT}{\mu m_p}
		\right]}\hspace{4cm}
	\nonumber
	\\
        &&= - RF\cos\theta.\hspace{1cm}
\end{eqnarray}
Rewriting equations (25), (26), and (27) in terms of the three dependent
variables, $v_\theta$, $v_\phi$, and $T$ we get,
\newcounter{subequation}[equation]
\renewcommand{\theequation}{\arabic{equation}\alph{subequation}}
\begin{eqnarray}
	\addtocounter{subequation}{+1}
	\lefteqn{v_\theta^2\frac{dv_\theta}{d\theta}
		+ \frac{4}{5}v_\theta^2\cos\theta
		\frac{d}{d\theta}
		\left[\frac{g_{\rm eff}}{g_{\rm eff}-g_{\rm rad}}
			\frac{kT/(\mu m_p)}{v_\theta\cos\theta}\right]}
	\hspace{3cm}
	\nonumber
	\\
	&=& -F_\theta-F_{\rm cen},
	\\
	\addtocounter{equation}{-1}
        \addtocounter{subequation}{+2}
	v_\theta v_\phi\frac{dv_\phi}{d\theta} &=& -F_\phi+F_{\rm cen},
	\\
	\addtocounter{equation}{-1}
        \addtocounter{subequation}{+3}
	\lefteqn{v_\theta^2\frac{dv_\theta}{d\theta} + \frac{2}{5}v_\theta
		\frac{d}{d\theta}
		\left[\frac{7g_{\rm eff}+3g_{\rm rad}}{g_{\rm eff}-g_{\rm rad}}
                        \frac{kT}{\mu m_p}\right]}
	\hspace{3cm}
	\nonumber
	\\
	&=& F_\phi-F_{\rm cen}-E_{\rm rad},
\end{eqnarray}
where the terms appearing on the right sides are:
the frictional dissipations in the $\phi$- and
$\theta$-directions,
\begin{eqnarray}
	F_\theta \equiv \alpha\frac{R}{h}
		v_\theta^2\sqrt{v_\theta^2+v_\phi^2},
	\\
	F_\phi \equiv \alpha\frac{R}{h}
                v_\phi^2\sqrt{v_\theta^2+v_\phi^2},
\end{eqnarray}
where $h=P/(\rho g_{\rm eff})=kT/[\mu m_p(g_{\rm eff}-g_{\rm rad})]$
is the pressure scale height,
the centrifugal term,
\begin{eqnarray}
	F_{\rm cen} \equiv v_\theta v_\phi^2\tan\theta, 
\end{eqnarray}
and the radiative dissipation
\begin{eqnarray}
        E_{\rm rad} \equiv \frac{RF}{y} = Rg_{\rm rad}v_\theta
		\frac{\dot{M}_{\rm Edd}}{\dot{M}}\cos\theta,
\end{eqnarray}
where $\dot{M}_{\rm Edd}\equiv4\pi Rc/\kappa$ is
the Eddington limit (with no rotation). In Appendix B we show how to solve
equations (28a)--(28c) for each of the derivatives of the dependent
variables to assist numerical integration.

\section{Solving for the Spreading Layer Properties}

  We now solve equations (28a)--(28c) by integrating from the equator toward
the pole and halting integration once the sink has been reached. There are
three boundary conditions at the equator for each of the dependent variables:
$v_{\theta,0}$, $v_{\phi,0}$, and $T_0$. Furthermore, we need an initial
angle at which to begin our integrations, $\theta_0$. In \S 4.1 we consider
the range of values appropriate for each of these variables, and show that
because the accretion rates are so much less than the Eddington limit,
the integrations naturally attract toward solutions where local viscous
dissipation is approximately balanced by local radiative cooling.
In \S 4.2 we analytically consider the differential equations to gain some
intuition before considering the numerical results, including showing how
$\theta_{\rm SL}$ changes with $\dot{M}$ and $\alpha$. In \S 4.3 we
present the full numerical solutions with descriptions of the important
SL properties.

\subsection{Boundary Conditions at the Equator}

  We begin by assuming that the accretion disk can be approximated by a
standard, thin Shakura-Sunyaev disk. Using the disk height from
Shakura \& Sunyaev (1973; in the limit of gas pressure much greater than
radiation pressure and using Kramer's opacity) and dividing by $R$ we find
the disk subtends an angle at the WD surface of
\begin{eqnarray}
	\theta_{\rm disk} = 1.5\times10^{-2}
	\alpha_{\rm disk}^{-1/10}
	\dot{M}_{18}^{3/20}
	M_1^{-3/8}
	R_9^{1/8},
\end{eqnarray}
where $\alpha_{\rm disk}$ is the viscosity parameter for the accretion disk
expected when the disk is actively accreting,
$\dot{M}_{18}\equiv\dot{M}/(10^{18} \textrm{ g s}^{-1})$,
$M_1\equiv M/M_\odot$, $R_9\equiv R/(10^9 \textrm{ cm})$,
and we set the factor $[1-(r/R)^{-1/2}]\approx1$.
We have chosen a rather high accretion rate for normalization because this
is the regime where the SL will have the most observational impact. The
material which comes in from the disk and subsequently spreads over the star
must then start at an angle $\theta_0\lesssim\theta_{\rm disk}$.
We therefore use a starting angle for integrations of $\theta_0\approx10^{-3}$.
In cases where the spreading angle is comparable or less than
$\theta_{\rm disk}$ we say that spreading is negligible and most of the
dissipated energy is radiated back into the disk. For the initial
azimuthal velocity we use $v_{\phi,0}=0.99v_{\rm K}$ so that the accreted
material is moving in nearly Keplerian orbits when it reaches the star. This
leaves two parameters to determine, $v_{\theta,0}$ and $T_0$.

  The initial temperature of the SL is set using the disk's midplane
temperature close to the WD. This is found from Shakura \& Sunyaev (1973) to be
\begin{eqnarray}
	T_{\rm disk} = 3.6\times10^{5}\textrm{ K }
        \alpha_{\rm disk}^{-1/5}
        \dot{M}_{18}^{3/10}
        M_1^{1/4}
        R_9^{-3/4},
\end{eqnarray}
Finally, we must set the value of $v_{\theta,0}$. To see that there is
a preferred value for this, consider setting $v_{\phi,0}$, $T_0$, and
$\theta_0$ using the description above. We then perform a number of
integrations, keeping all boundary values fixed while slightly varying
$v_{\theta,0}$ each time. The results of these integrations are shown in
Figure 2. In the top panel we see that the $v_\theta$ is
naturally attracted toward a specific solution.

\begin{figure}
\epsscale{1.2}
\plotone{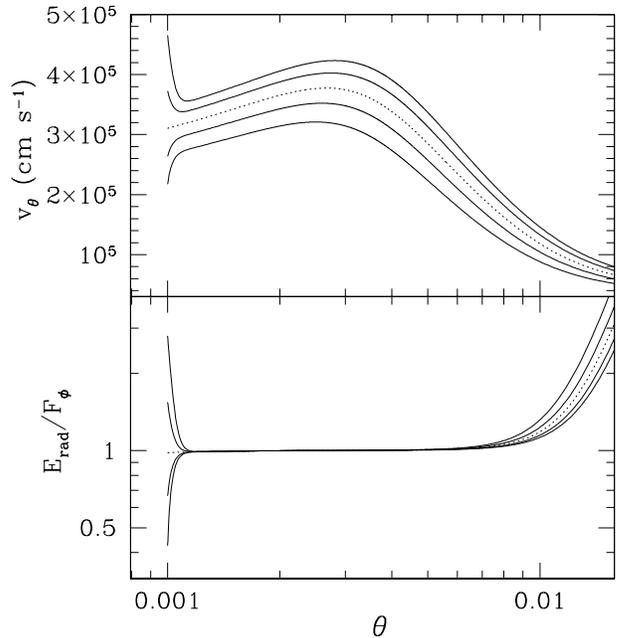}
\figcaption{A number of sample integrations used to constrain the boundary
condition at the equator. For all of the integrations we fix the values of
$\theta_0=10^{-3}$, $v_{\phi,0}=0.99v_{\rm K}$,
$\alpha=10^{-3}$, $\alpha_{\rm disk}=0.1$,
and $\dot{M}=3\times10^{17}\textrm{ g s}^{-1}$, for a white dwarf with
$M=0.6M_\odot$ and $R=9\times10^{8}\textrm{ cm}$. $T_0$ is fixed
using equations (34). Only $v_{\theta,0}$ is slightly different each time.
Looking at the top panel it is clear that there is one preferred solution for
$v_\theta$ (dotted line). The bottom panel shows the ratio $E_{\rm rad}/F_\phi$
for the same set of integrations. Comparing the two panels, we see that
this preferred solution corresponds to $E_{\rm rad}\approx F_\phi$.}
\end{figure}

  To find out the physical meaning behind this preferred solution, we consider
these same integrations, instead following the ratio $E_{\rm rad}/F_\phi$,
in the bottom panel of Figure 2. This ratio reflects how efficiently the
system is locally radiating the energy dissipated from friction. If
$E_{\rm rad}<F_\phi$ then there must be advection
because there is more local viscous heating than can be
radiated away. On the other hand, if $E_{\rm rad}>F_\phi$ the material
loses internal energy because it is radiating so efficiently. (We can
ignore $F_\theta$ because it is so much less than $F_\phi$.)
Comparing the top and bottom panels, we see that the asymptotic
solution found for $v_\theta$ appears to correspond to the case when initially
$E_{\rm rad}=F_\phi$, which is when all of the viscous dissipation
is being radiated away locally. To investigate further, we take
a closer look at $E_{\rm rad}/F_\phi$ in the top panel of Figure 3.
This shows that in fact the attracting solution is
$E_{\rm rad}\lesssim F_\phi$, which means that a small amount of
advection must take place. In the bottom panel of Figure 3 we
see that the actual ratio depends on the accretion rate and that the
higher the accretion rate is, the more advection takes place.
With this information in hand, we then use $E_{\rm rad}\lesssim F_\phi$
to get a closed set of equations so that the SL can be solved uniquely
for a given $\dot{M}$ and $\alpha$ (see Appendix C for additional details).

\begin{figure}
\epsscale{1.2}
\plotone{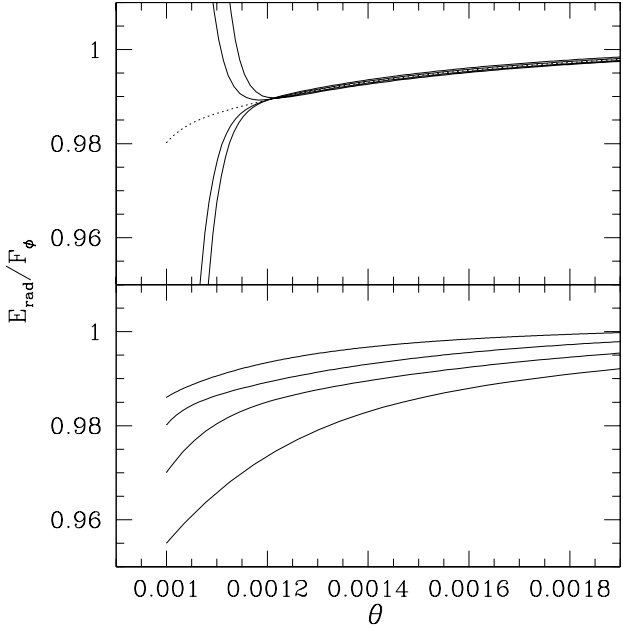}
\figcaption{The top panel shows the same integrations as in Figure 2,
but focusing near the equator. This shows that the actual
asymptotic solution has $E_{\rm rad}\lesssim F_\phi$, so that
there is naturally a small amount of advection taking place.
The bottom panel shows that solutions for accretion rates of
$10^{17}$, $3\times10^{17}$, $5\times10^{17}$, and $10^{18}\textrm{ g s}^{-1}$
(from top to bottom) with all other values the same as in Figure (2).
All solutions show $E_{\rm rad}\lesssim F_\phi$ 
and furthermore, this inequality increases with higher accretion rate.
As the accretion rate increases toward the Eddington limit we expect a
great deal of advection as found by Inogamov \& Sunyaev (1999) for the
neutron star case.}
\end{figure}

  This boundary condition is different than the NS case studied by IS99 where
it was found that the solutions are attracted toward the Eddington limit
(i.e., $g_{\rm eff}=g_{\rm rad}$) and there is a great deal of advection.
The differences between these results highlights the two limits of SL
solutions. In the case of NSs, the local accretion rate through the disk
is much closer to the Eddington rate making advection important. The low
$\dot{M}$ limit is the case of WDs considered here where radiation pressure
is negligible so that little advection takes place. A similar regime likely
exists for NSs at low accretion rates as well. In Figure 4 we show the initial
latitudinal velocity, $v_{\theta,0}$, using the condition $E_{\rm rad}=F_\phi$,
along with $g_{\rm eff}=g_{\rm rad}$. At low accretion
rates the two conditions are fairly different and the WD SL solutions are
shown to have little advection
and differ greatly from the NS case. As the accretion rate increases, the
two conditions asymptote together and the WD solutions become more like the
NS case. For boundary values at the right of the maximum in $v_{\theta,0}$
much more advection is seen in the solutions.

\begin{figure}
\epsscale{1.2}
\plotone{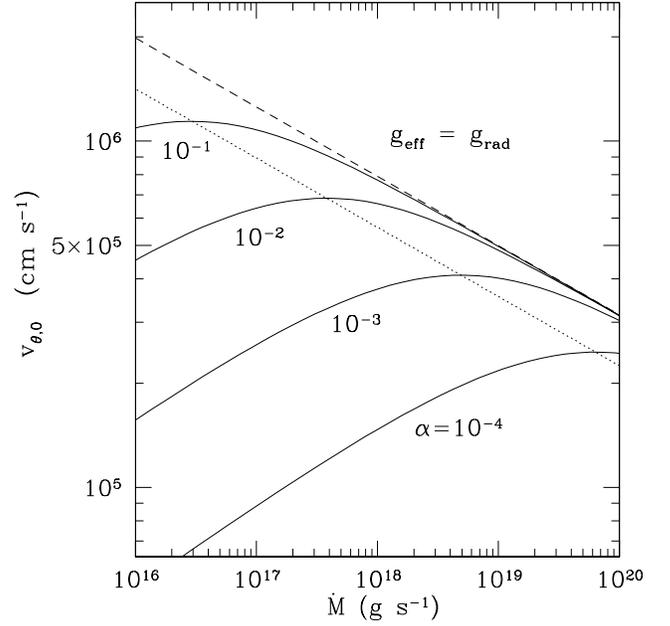}
\figcaption{The initial latitudinal velocity, $v_{\theta,0}$, versus accretion
rate, using the boundary condition that $E_{\rm rad}=F_\phi$ (solid lines).
The dashed line denotes the Eddington limit $g_{\rm eff}=g_{\rm rad}$
and the dotted lines pass through the maximum value $v_{\theta,0}$ can attain
for a given viscosity, $\alpha$. Solutions taken from the left of the dotted
line show little advection and a strong attraction toward the
$E_{\rm rad}=F_\phi$. Solutions using $v_{\theta,0}$ from the right of the
dotted line show much more advection, similar to the case of neutron
stars investigated by Inogamov \& Sunyaev (1999).}
\end{figure}

  Given the prescription described above there is still an additional
constraint for the values $\dot{M}$ and $\alpha$ to be consistent with
our plane parallel model which assumes that $h\ll R$.
Substituting the boundary condition that $E_{\rm rad} = F_\phi$ into
$h<R$ provides the constraint
\begin{eqnarray}
	\alpha\lesssim 10^{-2}
        \alpha_{\rm disk}^{3/5}
        \dot{M}_{18}^{-9/10}
        M_1^{-5/4}
        R_9^{7/4}.
\end{eqnarray}
This shows that at high $\dot{M}$, or when there is a lot of viscous
dissipation because of a high $\alpha$, it is possible that the SL will
expand so that $h\approx R$. This case is interesting for some high
$\dot{M}$ systems (such as symbiotic binaries which can have
$\dot{M}\approx10^{20}\textrm{ g s}^{-1}$), so although outside the scope
of this paper, such regions of parameter space should be studied
in the future.

\subsection{Analytic Estimates}

  As a first attempt at approximating the spreading angle, $\theta_{\rm SL}$,
one might guess that it is given by geostrophic balance (when the pressure
gradient in
the latitudinal direction is balanced by the Coriolis force pushing back toward
the equator). This can be found using conservation of momentum in the
$\theta$-direction, equation (25). Approximating $v_\theta\ll v_\phi$,
$g_{\rm rad}\ll g_{\rm eff}$, and $\theta\ll1$, this results in
\begin{eqnarray}
	\frac{d}{d\theta}\frac{kTy}{\mu m_p}
        \approx - yv_\phi^2\theta.
\end{eqnarray}
This then gives $\theta_{\rm SL}$ approximately as
\begin{eqnarray}
	\theta_{\rm SL}\approx\left(\frac{kT}{\mu m_p}\right)^{1/2}v_\phi^{-1}.
\end{eqnarray}
This may provide the correct estimate when friction is negligible, but it gives
the incorrect scalings for the case we are considering here where friction is
the primary agent for slowing the flow.

  Using this insight, we now derive approximate solutions, instead assuming
that conservation of $\phi$-momentum is controlling the flow. This will
illuminate the general features of the SL and provide power law scalings
consistent with numerical integrations. We first must estimate the size
and scaling of $v_\theta$. Setting $E_{\rm rad} = F_\phi$ near the equator
we find
\begin{eqnarray}
	Rg_{\rm rad}v_\theta
                \frac{\dot{M}_{\rm Edd}}{\dot{M}}\approx
			\alpha\frac{R}{h}
                	v_\phi^2\sqrt{v_\theta^2+v_\phi^2}
	\approx\alpha\frac{R}{h}v_\phi^3.
\end{eqnarray}
Solving equation (38) for $v_\theta$ gives the approximate latitudinal velocity
in the SL. Using equation (34) to set the temperature we find
\begin{eqnarray}
        v_{\theta,\rm SL}=2.1\times10^{6}\textrm{ cm s}^{-1}
	\alpha_{\rm disk}^{1/2}
	\alpha_{3}^{1/2}
        \dot{M}_{18}^{1/4}
        M_1^{5/8}
        R_9^{-7/8},
\end{eqnarray}
where $\alpha_{3}\equiv\alpha/10^{-3}$.

  To understand the effect of friction on setting $\theta_{\rm SL}$ we use
conservation of momentum in the $\phi$-direction,
equation (28b), which near the equator is
\begin{eqnarray}
	v_\theta v_\phi\frac{dv_\phi}{d\theta}=-F_\phi
		=-\alpha\frac{R}{h}v_\phi^3.
\end{eqnarray}
We can then estimate
\begin{eqnarray}
	\theta_{\rm SL}\approx\frac{h}{R}
		\frac{v_\theta v_\phi}{\alpha v_\phi^2}.
\end{eqnarray}
This gives the simple explanation that $\theta_{\rm SL}$ is set by two
factors: (1) the
pressure gradient in the latitudinal direction (the first term), and (2)
the competition between $\phi$-momentum fluxing the in the $\theta$-direction,
$v_\theta v_\phi$, and dissipation of $\phi$-momentum due to viscous stress,
$\alpha v_\phi^2$. This scaling is different than the results of IS99 who
found that the spreading flow adjusts itself so as to radiate everywhere
at close to the local Eddington rate, and thus their $\theta_{\rm SL}$
depends only on the ratio $\dot{M}/\dot{M}_{\rm Edd}$. At much less than the
Eddington
rate, equation (41) shows that $\theta_{\rm SL}$ is proportional to $h$ and
therefore highly dependent on the initial properties of the material when it
first reaches the WD equator. We can find the functional relationship between
$v_\phi$ and $\theta$ by assuming that $v_\theta$ and $T$ are constant and
integrating equation (40). Setting $g_{\rm eff} = GM/R^2-v_\phi^2/R$ and
using $\theta_0\approx0$ this results in
\begin{eqnarray}
	\theta = \frac{1}{\alpha}
		\frac{kT}{\mu m_p v_{\rm K}^2}
		\frac{v_\theta}{v_{\rm K}}\hspace{5cm}
		\nonumber
		\\
	\times\left( \frac{v_{\rm K}}{v_\phi}-\frac{v_{\rm K}}{v_{\phi,0}}
	+\log\frac{\sqrt{1-v_\phi/v_{\rm K}}}{\sqrt{1+v_\phi/v_{\rm K}}}
-\log\frac{\sqrt{1-v_{\phi,0}/v_{\rm K}}}{\sqrt{1+v_{\phi,0}/v_{\rm K}}}
		\right)
\end{eqnarray}
Figure 5 shows this result plotted alongside the numerical solutions found
in the next section. The analytic estimate shows only qualitative agreement,
which is not surprising given the crudeness of the estimates made, namely that
both $v_\theta$ and $T$ are constant. We estimate the spreading angle,
$\theta_{\rm SL}$, by assuming that $v_\phi\approx0.1v_{\rm K}$ at the
end of the spreading. In the limit of $v_{\phi,0}\approx v_{\rm K}$ we find
\begin{eqnarray}
	\theta_{\rm SL} = \frac{1}{\alpha}
                \frac{kT}{\mu m_p v_{\rm K}^2}
                \frac{v_\theta}{v_{\rm K}}
	\left( 9+\frac{1}{2}\log\frac{18/11}{1-v_{\phi,0}/v_{\rm K}}\right).
\end{eqnarray}
This result has a singularity at $v_{\phi,0}=v_{\rm K}$ (which it must because
material moving at the Keplerian velocity will not settle onto the WD), but
even for values $v_{\phi,0}\lesssim v_{\rm K}$ this last term is only of
order a few and has little effect on this angle estimate.
Substituting $v_\theta$ and $T$ by using equations (39) and (34), and setting
$v_{\phi,0}=0.99v_{\rm K}$ we find
\begin{eqnarray}
	\theta_{\rm SL} =
		2.4\times10^{-2}
		\alpha_{\rm disk}^{3/10}
		\alpha_{3}^{-1/2}
	        \dot{M}_{18}^{11/20}
        	M_1^{-5/8}
        	R_9^{-1/8}.
\end{eqnarray}
The scalings predicted by equation (44) are consistent with those found
performing the numerical integrations (as we describe in the following
section). This derivation shows that the change in $v_\phi$ and
the value of $\theta_{\rm SL}$ are mainly dictated by viscous dissipation.
Using $(d/dt)(yv_\phi)=-\alpha\rho v_\phi^2$, we estimate a timescale for
spreading of
\begin{eqnarray}
	t_{\rm SL} = h/(\alpha v_\phi)
	\approx50\textrm{ s }
	\alpha_{\rm disk}^{-1/5}
	\alpha_{3}^{-1}
        \dot{M}_{18}^{3/10}
        M_1^{-5/4}
        R_9^{7/4}.
\end{eqnarray}
This timescale is extremely short and mainly due to the small amount of
material which is in the SL at any given time. This implies that
if accretion decreases dramatically (such as following a dwarf nova
outburst) the hot belt around the equator will spread fairly
quickly over the WD surface and not be visible days or even hours later.

\begin{figure}
\epsscale{1.2}
\plotone{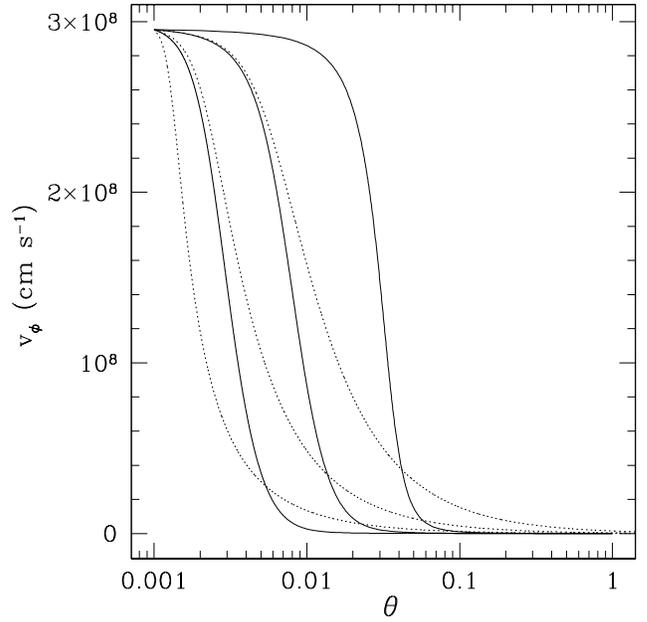}
\figcaption{A comparison between the numerical integrations (solid lines)
with the analytic estimate given by equation (42) (dotted lines). Accretion
rates of $\dot{M}=10^{17}$, $10^{18}$, and $10^{19} \textrm{ g s}^{-1}$
are shown (from left to right), all with $\alpha=10^{-3}$,
$\alpha_{\rm disk}=0.1$, $M=0.6M_\odot$, and $R=9\times10^{8}\textrm{ cm}$.
Though quantitatively different, the two results are qualitatively similar
and both predict similar spreading angles at
$v_\phi=0.1v_{\rm K}\approx3\times10^7\textrm{ cm s}^{-1}$.
At higher accretion
rates the solutions differ more because the analytic estimate ignores
radiation pressure effects.}
\end{figure}

  Since at low accretion rates $\theta_{\rm SL}\lesssim\theta_{\rm disk}$,
the spreading material will most likely be obscured by the disk,
making direct observations of the SL difficult. Setting
$\theta_{\rm SL}=\theta_{\rm disk}$ we solve for the critical $\dot{M}$
at which the SL will extend above the disk,
\begin{eqnarray}
	\dot{M}_{\rm crit} > 3.7\times10^{17} \textrm{ g s}^{-1}
	\alpha_{\rm disk}^{3/2}
	\alpha_{3}^{5/4}
        M_1^{15/8} 
        R_9^{5/8}.
\end{eqnarray}
The large value of this accretion rate implies that the effects of the SL may
only be seen in a select number of systems where a high $\dot{M}$ occurs such
as dwarf novae, symbiotic binaries, and supersoft sources. Lower accretion
rate sources must still have dissipation occuring at the interface between
the disk and WD surface, but the radiation created in these cases is most
likely absorbed and re-radiated by the accretion disk. We discuss all of
these issues in more detail in \S 5.

  Using the results from above
we find that the local flux in the SL is approximately,
\begin{eqnarray}
	F_{\rm SL}= 1\times10^{17}\textrm{ erg cm}^{-2}\textrm{ s}^{-1}
	\alpha_{\rm disk}^{-3/10}
        \alpha_{3}^{1/2}
	\dot{M}_{18}^{9/20}
        M_1^{13/8}
        R_9^{-23/8}.
\end{eqnarray}
Multiplying by the area of the spreading zone which is
$4\pi R^2\sin\theta_{\rm SL}\approx4\pi R^2\theta_{\rm SL}$
provides the luminosity. Comparing equations (44) and (47) we find
that $L\propto\dot{M}$ and independent of both $\alpha$ and $\alpha_{\rm disk}$
as it must be since these are free parameters. The value of the
luminosity is consistent with standard energy conservation
arguments, $L=GM\dot{M}/(2R)$. This means that luminosity alone will not
tell us whether a SL is present. A better test would be a study of
the emitting area and $T_{\rm eff}$. The emitting area scales like
$\theta_{\rm SL}$ which is expressed above. The $T_{\rm eff}$ is
given by the flux so that from equation (47) we find
\begin{eqnarray}
       T_{\rm eff, SL}
	= 2\times10^{5}\textrm{ K }
	\alpha_{\rm disk}^{-3/40}
        \alpha_{3}^{1/8}
        \dot{M}_{18}^{9/80}
        M_1^{13/32}
        R_9^{-23/32}.
\end{eqnarray}
This value is not so different than what would have been approximated from
BL arguments (for example see Frank, King \& Raine 2002).
This is not surprising since even if the SL increases
the emitting area by a factor of 10, the effective temperature would
only change by a factor of 1.8. The best test to see if spreading is
taking place is to follow how $T_{\rm eff}$ changes with $\dot{M}$.
The SL model shows that $T_{\rm eff}$ will change much more gradually
than if the radiating area on the WD is fixed (which would predict
$T_{\rm eff}\propto\dot{M}^{1/4}$).

\subsection{Numerical Solutions}
  
  We now present the full numerical solutions of equations (28a)--(28c).
These confirm the general estimates made in the previous section and also
provide additional details that cannot be investigated analytically.
For these integrations we use the fiducial values of $M=0.6M_\odot$,
$R=9\times10^{8}\textrm{ cm}$, and $\alpha_{\rm disk}=0.1$. The composition
of the accreting material is assumed to be solar so that $\mu=0.62$ and
$\kappa=0.34\textrm{ cm}^{2}\textrm{g}^{-1}$.

\begin{figure}
\epsscale{1.2}
\plotone{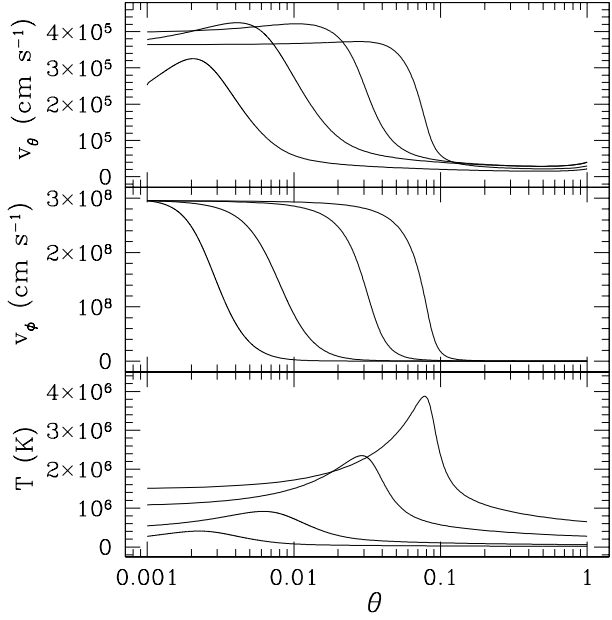}
\figcaption{The velocities and temperature of the spreading layer
for a white dwarf with $M=0.6M_\odot$ and $R=9\times10^8\textrm{ cm}$,
with $\alpha=10^{-3}$ and $\alpha_{\rm disk}=0.1$. From left to right
the accretion rates are $10^{17}$, $10^{18}$, $10^{19}$ and
$3\times10^{19}\textrm{ g s}^{-1}$.}
\end{figure}

\begin{figure}
\epsscale{1.2} 
\plotone{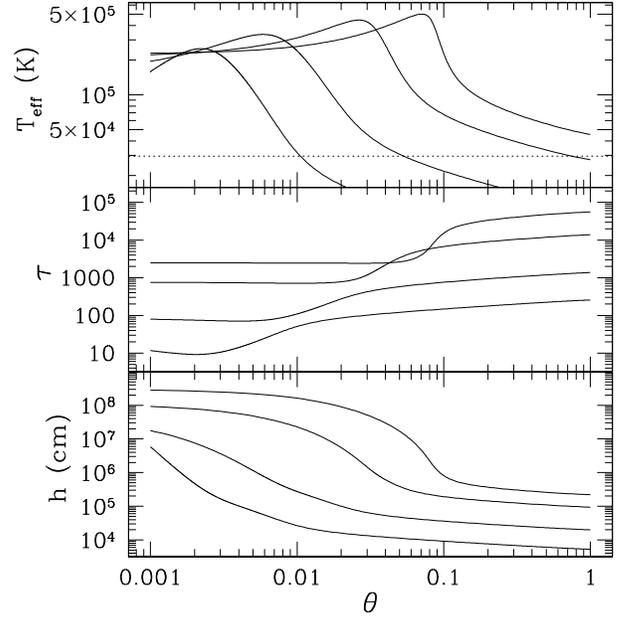}
\figcaption{The effective temperature, optical depth, and pressure scale 
height of the spreading layer, using the same parameters as in Figure 6.
The dotted line shows $T_{\rm eff}=3\times10^4\textrm{ K}$, a fiducial
temperature for the underlying accreting white dwarf.}
\end{figure}

\begin{figure}
\epsscale{1.2}
\plotone{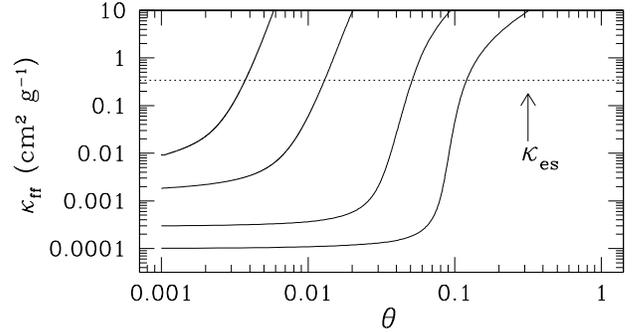}
\figcaption{The free-free opacity, $\kappa_{\rm ff}$, 
calculated using the properties of the
spreading layer solutions, using the same parameters as in Figure 6.
This can be compared to the electron scattering opacity,
$\kappa_{\rm es}=0.34\textrm{ cm}^2\textrm{ g}^{-1}$
(dotted line), to see that the opacity is correctly approximated for the
majority of the spreading layer, especially in the regions which provide
the majority of the flux.}
\end{figure}

  First we set our boundary value for $T_0$ using the temperature
of equation (34), and then we set $v_{\theta,0}$ by using the condition that
$E_{\rm rad}\lesssim F_\phi$. The integrations are then performed on the three
coupled differential equations as derived in Appendix B from equations
(28a)--(28c). In Figures 6 and 7 we plot the properties of the SL for
$\alpha=10^{-3}$ and a range of accretion rates. At low accretion rates
($\dot{M}\approx10^{17}-10^{18}\textrm{ gs}^{-1}$) we find the numerical
integrations result in values close to what was approximated analytically
in the previous section. At the higher accretion rates
($\dot{M}\gtrsim10^{19}\textrm{ gs}^{-1}$) more advection takes place,
resulting in solutions which look similar to what IS99 found for
neutron stars. Most notably, the temperature profiles show large peaks
at high latitudes due to energy being advected away from the equator and
deposited higher up on the star. In Figure 7 we denote the fiducial
$T_{\rm eff}$ expected for the underlying accreting WD. Comparing this and
Figure 6 shows that once the SL has come into thermal balance with the
underlying WD, the SL is also rotating with the WD ($v_\phi\approx0$).
The optical depth is shown to always be much greater than unity, so that
our assumptions of a radiation pressure acting in our one-zone envelope
and a true $T_{\rm eff}$ are valid. At high accretion rates the scale
height becomes comparable to $R$, showing the limit of the plane parallel
model. Our viscous stress parametrization becomes questionable at low
accretion rates ($\dot{M}\lesssim10^{17}\textrm{ g s}^{-1}$) when the
latitudinal shear is much greater than the shear between the SL and
underlying WD (because $R\theta_{\rm SL}<h$). Qualitatively, this does
not change the important result that most or all of the SL dissipation
happens within the width of the accretion disk at these accretion rates.
Figure 8 shows the free-free opacity, calculated from the solutions
that are found using a constant electron scattering opacity. For all
solutions, the free-free opacity is small in the hot portions of the SL.
This gives some confidence that $T_{\rm eff}$
over this region is accurate, because the majority of the flux is radiated
here. Unfortunately, our opacity assumption always breaks
down in regions where the SL is cool which makes any spectrum calculation
based on this model somewhat dubious. In Figures 9 and 10 we show that the
properties of the SL can vary depending on the value of the viscosity.

\begin{figure}
\epsscale{1.2}
\plotone{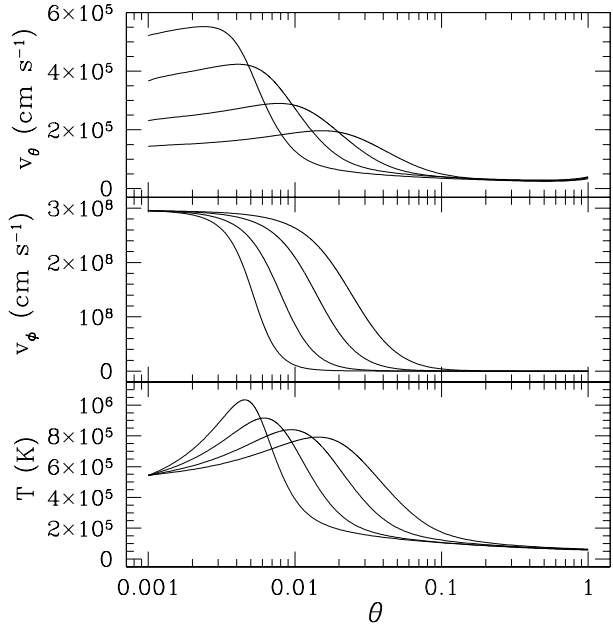}
\figcaption{The velocities and temperature of the spreading layer
for a white dwarf with $M=0.6M_\odot$ and $R=9\times10^8\textrm{ cm}$,
with $\dot{M}=10^{18}\textrm{ g s}^{-1}$ and $\alpha_{\rm disk}=0.1$.
From left to right the values of $\alpha$ are $3\times10^{-3}$,
$10^{-3}$, $3\times10^{-4}$, and $10^{-4}$.}
\end{figure}

\begin{figure}
\epsscale{1.2}
\plotone{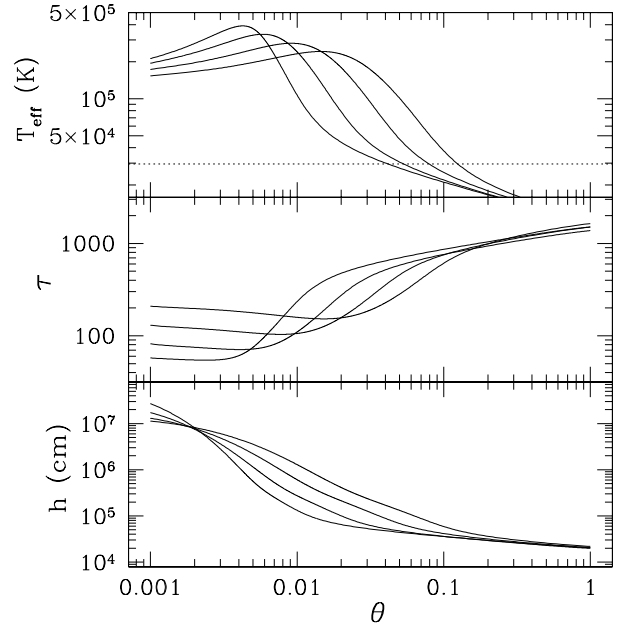}
\figcaption{The effective temperature, optical depth, and pressure scale
height of the spreading layer, using the same parameters as in Figure 9.
Viscous stress decreases from left to right.
The dotted line shows $T_{\rm eff}=3\times10^4\textrm{ K}$, a fiducial
temperature for the underlying accreting white dwarf.}
\end{figure}

  Figure 11 summarizes the key feature of the SL which is how the spreading
angle (and thus the radiating area) changes with accretion rate. At low
accretion rates the scaling is consistent with what was found analytically,
equation (44). Figure 11 reinforces the result that high accretion rates
are needed for any hope of appreciable spreading to take place.
It is important to note that even when the accretion disk covers the main
portion of the SL, the SL may still have some observational impact depending
on the viewing angle of the WD and how far the contrast in
$T_{\rm eff}$ extends (see Figures 7 and 10).

\begin{figure}
\epsscale{1.4}
\plotone{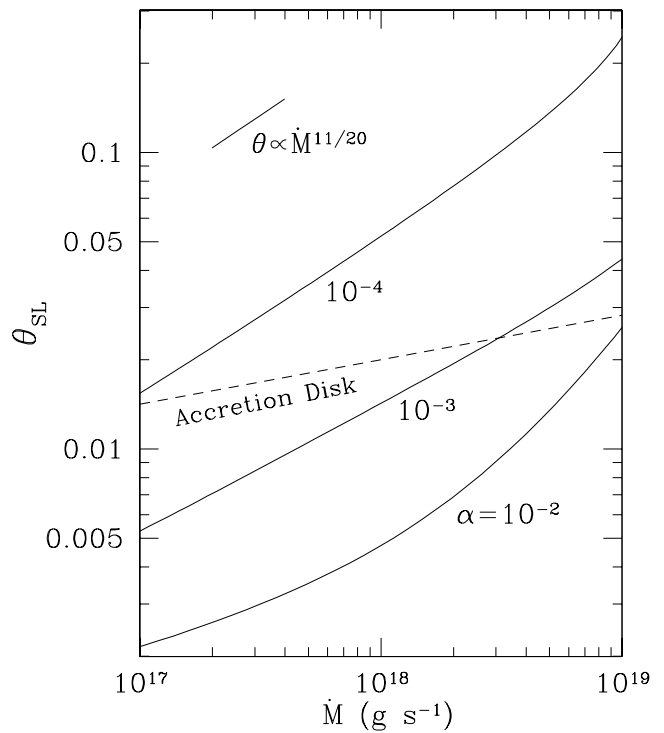}
\figcaption{The spreading angle, $\theta_{\rm SL}$, versus accretion
rate for a white dwarf with $M=0.6M_\odot$ and $R=9\times10^9\textrm{ cm}$
and three different viscous stress parameters. The spreading angle is
defined as the angle at which $v_\phi=0.1v_{\rm K}$.
At low accretion rates the scaling with $\dot{M}$ is
fairly consistent with what is expected from an analytic
approximation which gives $\theta_{\rm SL}\propto\dot{M}^{11/20}$.
This also shows that high accretion rates are needed for the
spreading layer to extend beyond the angle made by a Shakura-Sunyaev disk,
equation (33) (dashed line).}
\end{figure}

\section{Initial Comparisons to Observations}

  From the analysis in \S 4, it is qualitatively clear that the SL is best
observed in high accretion rate systems so that it is not obscured by the
accretion disk or its winds. This suggests that the best candidates to
show a SL are dwarf novae, symbiotic binaries, and supersoft
sources. We now review the current observations of these accreting WDs in
the EUV and soft X-rays.

\subsection{Dwarf Nova Outbursts}

  There have been a number of observations of dwarf novae in outburst or
superoutburst using the {\it Extreme Ultraviolet Explorer} ({\it EUVE}),
including those of SS Cyg
(Mauche, Raymond \& Mattei 1995; Wheatley, Mauche \& Mattei 2003,
hereafter WMM03), U Gem (Long et al. 1996), VW Hyi (Mauche 1996) and OY Car
(Mauche \& Raymond 2000). Mauche, Mattei \& Bateson (2001) provides a review
of many of these objects.

  WMM03 provide an especially interesting observation of an outburst of SS
Cygni studied in the optical by The American Association of Variable Star
Observers (AAVSO), in the X-rays by the {\it Rossi X-ray Timing Explorer}
({\it RXTE}), and in the EUV using {\it EUVE}. The observations show a
complete optical light curve, following the burst over $\sim12\textrm{ days}$.
The X-rays show two small bursts of activity, the first coming
$\sim1.5\textrm{ days}$ after the initial rise
in the optical and the second coming near the end of the outburst. In
between there is emission in the EUV. The correlations between these
three bands are consistent with what is expected by the standard dwarf nova
picture. The outburst begins once the disk is in the high state, starting
the rise in optical, and then proceeds from the outside of the disk in.
Once material begins accreting onto the WD it is initially optically thin,
and thus radiates in the X-rays at a temperature given by the virial
temperature of the accreting gas (Pringle \& Savonije 1979; Tylenda 1981;
Narayan \& Popham 1993). This provides the initial $\sim1.5\textrm{ days}$
of X-rays. Once the material becomes optically thick, it radiates in the EUV
for most of the remainder of the burst (Pringle 1977; Tylenda 1977;
Popham \& Narayan 1995). Finally, at the end, the material becomes optically
thin one last time, radiating a little more in the X-rays before the outburst
ceases.

  Qualitatively, a similar picture is also expected for a SL, even though
the calculations we perform above only apply when the SL is optically
thick. One might expect the SL to be important because of the large inferred
accretion rates at the peak of the outburst ($\sim10^{18}\textrm{ g s}^{-1}$).
As discussed in \S 4.2, evidence that spreading is occuring requires studying
the change in effective temperature with accretion rate. If the radiating area
does not change with accretion rate then one expects
$T_{\rm eff}\propto\dot{M}^{1/4}$. On the other hand, since the area increases
with $\dot{M}$ for the SL we find $T_{\rm eff}\propto\dot{M}^{9/80}$,
equation (48). Figure 12 shows the count rate (which is
proportional to luminosity) versus the hardness ratio (related to
$T_{\rm eff}$) taken from this observation by WMM03. Fit are the two power
laws expected from a SL and a constant radiating area. Even
though this does not show conclusively that the SL exists, this shows in
principle the kind of test which needs to be done as data improves.
Other observations of SS Cyg (such as Mauche, Raymond \& Mattei 1995), also
report a fairly constant hardness ratio during outburst even though the
count rate changes by two orders of magnitude.

\begin{figure}
\epsscale{1.2}
\plotone{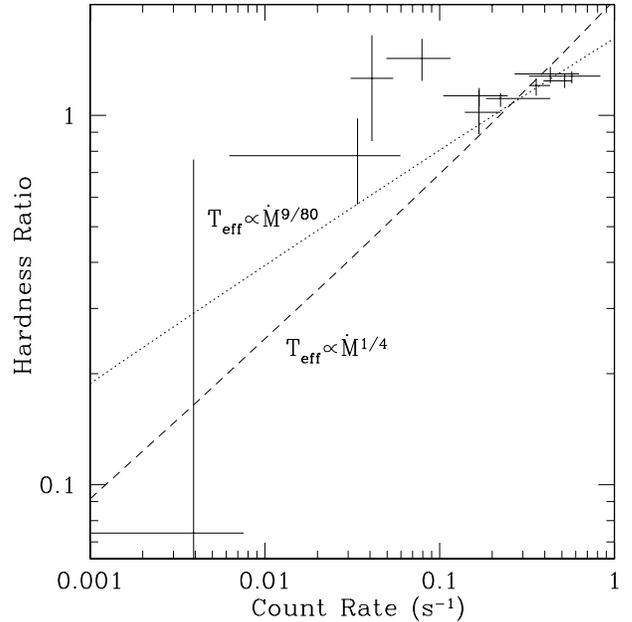}
\figcaption{Hardness ratio versus count rate as measured during
a dwarf nova outburst of SS Cyg by Wheatley, Mauche \& Mattei (2003).
The hardness ratio is defined as the ratio of counts between the
the $72-95\textrm{ \AA}$ band (excluding $76-80\textrm{ \AA}$)
to the $95-130\textrm{ \AA}$ band. Plotted over the
data are the relations expected from the spreading layer (dotted line)
and a model which has constant radiating area (dashed line).
In each case the area is assumed to radiate like a blackbody
and the Wein limit is used. The scaling between count rate and accretion
rate is treated as a free parameter used to maximize the fit in each case.
The spreading layer gives a shallower power law, as expected because
its radiating area increases with accretion rate, but it provides only
a slightly better fit to the data.}
\end{figure}

  Another suggestive result by Mauche, Raymond \& Mattei (1995) is the
count rate in the EUV as compared with typical count rates in the optical
for similar outbursts. If the former is assumed to represent the region up
close to the WD and the latter the accretion disk, these observations
suggest that the ratio of the two luminosities is
$L_{\rm BL}/L_{\rm disk}\lesssim0.07$. This is in stark contrast from
$L_{\rm BL}/L_{\rm disk}\sim1$ as expected from BL models. It is possible
that because the SL only extends a small distance above the disk, much of
the radiation from the SL is obscured by, or contained within, the
accretion disk or its winds.

  A much more complicated picture is provided by the superoutburst of the
edge-on binary OY Car (Mauche \& Raymond 2000). By comparing optical and
EUV lightcurves during outburst they find that
while the optical is eclipsed, the EUV shows no sign of being eclipsed.
This calls into question the assumption that the EUV is associated with
a region near the WD and suggests that perhaps the EUV is from a much
more extended region than previously thought. Good fits are made to
the outburst spectrum by assuming that the radiation from both the BL
and accretion disk are scattered into the observer's line of sight by
an ionized accretion disk wind. These additional complications make it
difficult to simply compare our SL with this or similar observation.

  In a study of U Gem using the \textit{Far Ultraviolet Spectroscopic Explorer}
(\textit{FUSE}), Froning et al. (2001) model the spectra both during outburst
and in the decline afterwards in the far-ultraviolet. They find that the
majority of the spectra during outburst is consistent with a steady state
accretion disk with $\dot{M}=4.6\times10^{17}\textrm{ g s}^{-1}$.
This is close to the critical accretion rate found in equation (46) so that
we might guess a SL is present. The accretion disk model underpredicts the
observed spectra at short wavelengths ($<950$\AA), which is a sign of a
missing hot component, perhaps the SL. Though suggestive, there is still
more work to be done to determine whether a SL has been seen. It is not
clear whether accretion disk model improvements would explain this difference.
In the same study by Froning et al. (2001), they also look at U Gem
a few days after the outburst peak when the luminosity is declining. They
find that the spectra are very well fit by a WD model. They also attempt a fit
which includes an additional hot belt on the WD, but they find that the
improvement of the fit is negligible. This is long after the timescale we
find for spreading, equation (45), so we do not expect to see a SL in this
case.

  In an observation of VW Hyi, Sion et al. (1996) use the \textit{Hubble}
Faint Object Spectrograph (FOS) to look at spectra a day after a dwarf nova
outburst and ten days after a super outburst. In each case they compare
spectral fits of a single WD model with the fits of a WD plus a hot belt,
and for both observations they find that the latter model provides a better
fit. From the solutions found for the SL in \S 4 it is surprising that
the hot belt is still present on the WD this many days after
outburst when the SL should have already spread over the star due to
friction. Other observations by the same group (Huang et al. 1996a, 1996b)
show similar, but less conclusive results.

\subsection{Symbiotic Binaries and Supersoft Sources}

  Another promising candidate for showing a SL are symbiotic binaries.
These are WDs which are accreting from the winds of red giants with
$\dot{M}=10^{17}-10^{20}\textrm{ g s}^{-1}$, and they typically show a hot
component which may or may not be coming from near the WD.
M\"{u}rset, Nussbaumer, Schmid \& Vogel (1991) review the characteristics of
this hot component for a large number of systems, finding luminosities in the
range $0.3 L_\odot\lesssim L\lesssim3.7\times10^4 L_\odot$ and effective
temperatures around
$5.5\times10^4\textrm{ K}\lesssim T_{\rm eff}\lesssim2.1\times10^5\textrm{ K}$.
From these two quantities the radius associated with the emitting region
can be estimated. This is typically $\sim10^{10}\textrm{ cm}$, much
larger than expected if the radiation where coming directly from the WD
surface, which suggests the presence of a hot ionized nebula surrounding
the WD. There are three systems listed in M\"{u}rset et al. (1991) which
have features qualitatively different than most other symbiotic binaries;
these are EG And, AG Dra, and CH Cyg. All three of this systems are
underluminous in comparison with the other symbiotic binaries which implies a
much smaller radius, closer to what is expected for WD. A May 1988 observation
of CH Cyg resulted in an inferred radius of $1.4\times10^8\textrm{ cm}$
which is even too small to be a WD. Though M\"{u}rset et al. (1991) argue
this could be due to an incorrect distance measurement, it could also be
a sign of the SL. Two of these systems show flickering in the optical
(EG And and CH Cyg; Sokoloski, Bildsten \& Ho 2001), which is a trait
uncommon to symbiotics, but common to accreting CVs. This may point to
disk accretion being present and thus also a SL.

  Finally, supersoft sources are WD binaries with
$\dot{M}\sim10^{18}-10^{20}\textrm{ g s}^{-1}$.
At these high accretion rates accreted hydrogen on the WD surface is believed
to be burning steadily (van den Heuvel et al. 1992; Heise, van Teeseling
\& Kahabka 1994) and this provides the large luminosity seem from these
sources ($\sim10^{37}-10^{38}\textrm{ erg s}^{-1}$). The SL would appear
at a significantly different $T_{\rm eff}$ from the WD surface because of
the difference in emitting area. The problem with these systems is that
the burning causes such high luminosities that
the SL would be dwarfed in comparison. Current observations of supersoft
sources do not have enough signal to show a SL and would require
much more detailed models to correctly determine whether a SL is present.

\section{Discussion and Conclusions}

  The SL model, first conceived by IS99, provides a new way of understanding
the properties of newly accreted material and its interaction with the
stellar surface. In this model the latitudinal direction is used as the
independent variable, as opposed to standard BL models which follow how
the disk changes near the star along the radial direction. This allows
an investigation of the radiating area of the hot belt which forms near
the equator, and it describes the properties of the accreting material when
it first comes into hydrostatic balance on the star. We have investigated the
solutions of the SL model when applied to WDs, accreting in the range of
$\dot{M}\sim10^{17}-10^{19}\textrm{ g s}^{-1}$. We find that the integrations
for the spreading flow naturally attract toward solutions in which initially
local radiative cooling balances local viscous dissipation.
As the accretion rate gets higher, we find increasingly more advection in the
flow, similar to the solutions for NSs investigated by IS99. The spreading
angle on WDs is found to typically be $\theta_{\rm SL}\approx0.01-0.1$,
depending on both the accretion rate and the viscosity. The SL
has an effective temperatures in the range of $(2-5)\times10^5\textrm{ K}$
with a pressure scale height of $10^7-10^8\textrm{ cm}$.

  To be clearly observed in actively accreting systems, the SL must extend to
a large enough angle to be seen above the accretion disk. This requires the
accretion rate to be high, $\dot{M}\gtrsim10^{18}\textrm{ g s}^{-1}$,
assuming that the accretion disk is similar to a thin Shakura-Sunyaev
disk. The best candidates to show the SL are dwarf novae in outburst,
supersoft sources, and symbiotic binaries. Current observations of the dwarf
novae may be the best opportunity to see if spreading is actually occurring.
The scaling of effective temperature with accretion rate is much weaker
in the case of a SL because of the change in radiating area. At lower accretion
rates, $\dot{M}\lesssim10^{18}\textrm{ g s}^{-1}$, the SL will not spread
far from the equator and most of the dissipated energy will radiate back into
the accretion disk. Although this may make comparison between theory and
observations difficult, because the SL cannot be seen directly, it may still
have important consequences for the accretion disk structure and spectra.

  The short timescale for spreading, $t_{\rm SL}\sim50\textrm{ s}$,
suggests that a hot belt around the WD equator will be difficult to see
after an outburst because the hot material should spread over
the WD surface quickly. This also means that mixing
between freshly accreted material and much deeper layers appears difficult on
accreting WDs. Rosner et al. (2001) and Alexakis et al. (2003) have argued
that the mixing of recently accreted H/He with the underlying layer of
C/O due to rotational shearing explains observations of CNO nuclei in the
ejecta of nova. To have $t_{\rm SL}$ of order the accretion time
($\sim10^3\textrm{ yr}$) requires $\alpha\lesssim10^{-11}$, so that the
viscosity must be even less than what is estimated from microphysics
(see Appendix A).
This makes it difficult to imagine mixing via this mechanism
all the way down to the C/O layers.

  The initial calculations by IS99 and in this paper describe some of the
general features of the SL model, a new area of investigation which will
lead to further studies of how accreted material settles onto stars. This
may include studying how differential rotation and angular momentum transfer
affect the underlying stellar surface or the possibility of nonradial
oscillations present in the SL. Spectral modeling of the SL, along
comparisons with observations, may also help in identifying if and when
spreading is present.

 We thank Rashid Sunyaev for helpful and enthusiastic discussions about
spreading, and Christopher Mauche for answering our questions about cataclysmic
variable observations. We have also benifitted from conversations with Phil
Arras, Philip Chang, Chris Deloye, Aristotle Socrates, and Dean Townsley.
We thank the referee for a careful reading of this paper along with
many helpful suggestions. This work was supported by the National Science
Foundation under grants PHY99-07949 and AST02-05956 and by the Joint
Institute for Nuclear Astrophysics through NSF grant PHY02-16783.

\begin{appendix}

\section{Viscous Stress Parametrization}

  In the derivation of the spreading differential equations we choose to
parametrize $\tau$ in terms of a unitless constant, $\alpha$, so that
\begin{eqnarray}
        \tau = \alpha \rho v^2.
\end{eqnarray}
We follow IS99 and estimate the
values of $\alpha$ implied by possible ionic or radiative viscosities.
In general the viscous stress is expected to be set by turbulence, so
these estimates only provide a lower limit to $\alpha$. The ion
viscosity, $\nu_i$, is related to the viscous stress by
\begin{eqnarray}
        \tau = \nu_i \rho \frac{\partial v}{\partial z}
	\approx \nu_i \rho \frac{v}{h}.
\end{eqnarray}
Using the
ion viscosity of Spitzer (1965) we find
$\nu_i\approx2\times10^6\textrm{ cm}^2\textrm{ s}^{-1}$
(using typical values of the spreading layer solutions from \S 4).
Equations (A1) and (A2) can be combined to find
$\alpha\sim10^{-10}$, which shows that this contribution
can be ignored. We next consider the radiative viscosity,
$\nu_r$, which is given by
\begin{eqnarray}
        \tau = \nu_r \rho_r \frac{\partial v}{\partial z}
        \approx \nu_r \rho_r \frac{v}{h},
\end{eqnarray}
where $\rho_r=aT^4/c^2$ is the radiation density. The radiative viscosity
can be be written as $\nu_r=\lambda c$, where $\lambda$ is the photon
mean free path given by $\lambda=1/(\rho\kappa)$. We then solve for
$\alpha$ and find it can be written in terms of three ratios as
\begin{eqnarray}
	\alpha = \frac{\rho_r}{\rho}
		\frac{\lambda}{h}
		\frac{c}{v}
		\sim 10^{-6},
\end{eqnarray}
which is small due to $\rho_r/\rho\ll1$.
Though larger than the former value for $\alpha$, this quick check shows
that neither of these mechanisms are large enough to provide the viscosity
needed to slow down the spreading material. We therefore assume that
there is a turbulent viscosity which results in higher values of
$\alpha$ and leave this as a free parameter.
In IS99, this form of the viscous stress, equation (A1), is contrasted with the
form typically used for accretion disks ($\tau=\alpha c_s h$, where $c_s$
is the speed of
sound; Shakura \& Sunyaev 1973). The latter
prescription for the viscous stress describes the friction between adjacent
annuli of the layer and IS99 show that it is negligible as long as $h/R\ll1$
(so that most of the free energy is at the SL/WD boundary).

\section{Rewriting the Differential Equations for Numerical Integration}

  To solve equations (28a)--(28c) numerically we must solve for each of the
derivatives in terms of the dependent variables.
This involves expanding the messy derivatives in equations
(28a) and (28c). From equation (28a) we need
\begin{eqnarray}
	\frac{d}{d\theta}
                \left[\frac{g_{\rm eff}}{g_{\rm eff}-g_{\rm rad}}
                        \frac{kT/(\mu m_p)}{v_\theta\cos\theta}\right]
        = \frac{g_{\rm eff}}{g_{\rm eff}-g_{\rm rad}}
		\frac{1}{v_\theta\cos\theta}\frac{d}{d\theta}\frac{kT}{\mu m_p}
        - \frac{g_{\rm eff}}{g_{\rm eff}-g_{\rm rad}}
                \frac{kT/(\mu m_p)}{v_\theta^2\cos\theta}
		\frac{dv_\theta}{d\theta}
        + \frac{1}{g_{\rm eff}-g_{\rm rad}}
                \frac{kT/(\mu m_p)}{v_\theta\cos\theta}
		\frac{dg_{\rm eff}}{d\theta}
	\nonumber
	\\
        \mbox{}- \frac{g_{\rm eff}}{(g_{\rm eff}-g_{\rm rad})^2}
                \frac{kT/(\mu m_p)}{v_\theta\cos\theta}
		\frac{d}{d\theta}(g_{\rm eff}-g_{\rm rad})
        + \frac{g_{\rm eff}}{g_{\rm eff}-g_{\rm rad}}
		\frac{kT/(\mu m_p)}{v_\theta\cos\theta}\tan\theta.
\end{eqnarray}
The individual derivatives in this expression are given by
\begin{eqnarray}
	\frac{dg_{\rm eff}}{d\theta}
	= -\frac{2v_\theta}{R}\frac{dv_\theta}{d\theta}
		\mbox{}- \frac{2v_\phi}{R}\frac{dv_\phi}{d\theta},
\end{eqnarray}
and
\begin{eqnarray}
        \frac{dg_{\rm rad}}{d\theta}
                = \left(\frac{4}{T}\frac{dT}{d\theta}
                        +\frac{1}{v_\theta}\frac{dv_\theta}{d\theta}
                        - \tan\theta\right)g_{\rm rad}.
\end{eqnarray}
We can then rewrite equation (28a) as
\begin{eqnarray}
        \left[v_\theta^2
                -\frac{4}{5}\frac{g_{\rm eff}^2-2g_{\rm eff}g_{\rm rad}
                        -2g_{\rm rad}v_\theta^2/R}{(g_{\rm eff}-g_{\rm rad})^2}
			\frac{kT}{\mu m_p}\right]
                \frac{dv_\theta}{d\theta}
        +\frac{8}{5}\frac{g_{\rm rad}}{(g_{\rm eff}-g_{\rm rad})^2}
		\frac{v_\theta v_\phi}{R}\frac{kT}{\mu m_p}
		\frac{dv_\phi}{d\theta}
        +\left[\frac{4}{5}\frac{g_{\rm eff}^2
                        +3g_{\rm eff}g_{\rm rad}}{(g_{\rm eff}-g_{\rm rad})^2}
			v_\theta\right]
                \frac{d}{d\theta}\frac{kT}{\mu m_p}
	\nonumber
	\\
        = -F_\theta - F_{\rm cen}
        - \frac{4}{5}\frac{g_{\rm eff}^2-2g_{\rm eff}g_{\rm rad}}
		{(g_{\rm eff}-g_{\rm rad})^2}
                v_\theta \frac{kT}{\mu m_p}\tan\theta.
\end{eqnarray}
From equation (28c), we must expand
\begin{eqnarray}
	\frac{d}{d\theta}\left[
		\frac{7g_{\rm eff}+3g_{\rm rad}}
		{g_{\rm eff}-g_{\rm rad}}\frac{kT}{\mu m_p}\right]
	= \frac{d}{d\theta}\left[
		\frac{10g_{\rm eff}-3(g_{\rm eff}-g_{\rm rad})}
		{g_{\rm eff}-g_{\rm rad}}\frac{kT}{\mu m_p}\right]
	= \frac{10g_{\rm eff}-3(g_{\rm eff}-g_{\rm rad})}
		{g_{\rm eff}-g_{\rm rad}}\frac{d}{d\theta}\frac{kT}{\mu m_p}
	+ \frac{10}{g_{\rm eff}-g_{\rm rad}}\frac{kT}{\mu m_p}
		\frac{dg_{\rm eff}}{d\theta}
	\nonumber
	\\
	\mbox{}- \frac{10g_{\rm eff}}{(g_{\rm eff}-g_{\rm rad})^2}
		\frac{kT}{\mu m_p}\frac{d}{d\theta}(g_{\rm eff}-g_{\rm rad}).
\end{eqnarray}
Using equations (B2) and (B3) we can then rewrite equation (28c) as
\begin{eqnarray}
        \left[v_\theta^2
		+4\frac{g_{\rm eff}g_{\rm rad}+2g_{\rm rad}v_\theta^2/R}
			{(g_{\rm eff}-g_{\rm rad})^2}
                        \frac{kT}{\mu m_p}\right]
                \frac{dv_\theta}{d\theta}
        +\frac{8g_{\rm rad}}{(g_{\rm eff}-g_{\rm rad})^2}
		\frac{v_\theta v_\phi}{R}\frac{kT}{\mu m_p}
                \frac{dv_\phi}{d\theta}
        +\left[\frac{2}{5}\frac{7g_{\rm eff}^2+36g_{\rm eff}g_{\rm rad}
                -3g_{\rm rad}^2}{(g_{\rm eff}-g_{\rm rad})^2}v_\theta\right]
                \frac{d}{d\theta}\frac{kT}{\mu m_p}
        \nonumber
        \\
        = F_\phi -F_{\rm cen} - E_{\rm rad}
        +\frac{4g_{\rm eff}g_{\rm rad}}{(g_{\rm eff}-g_{\rm rad})^^2}
		v_\theta \frac{kT}{\mu m_p}\tan\theta.
\end{eqnarray}
Equations (B4), (28b), and (B6) are of the form
\begin{eqnarray}
        A_1\frac{dv_\theta}{d\theta}+A_2\frac{dv_\phi}{d\theta}
		+A_3\frac{d}{d\theta}\frac{kT}{\mu m_p}&=&C_1,
        \\
        v_\theta v_\phi\frac{dv_\phi}{d\theta}&=&C_2,
        \\
        B_1\frac{dv_\theta}{d\theta}+B_2\frac{dv_\phi}{d\theta}
		+B_3\frac{d}{d\theta}\frac{kT}{\mu m_p}
                &=&C_3.
\end{eqnarray}
This can be inverted to find,
\begin{eqnarray}
        \frac{dv_\theta}{d\theta}&=&\frac{B_3C_1-A_3C_3
			+(A_3B_2-A_2B_3)C_2/(v_\theta v_\phi)}
                {A_1B_3-A_3B_1},
        \\
        \frac{dv_\phi}{d\theta}&=&\frac{C_2}{v_\theta v_\phi},
        \\
        \frac{d}{d\theta}\frac{kT}{\mu m_p}&=&\frac{A_1C_3-B_1C_1
			+(A_2B_1-A_1B_2)C_2/(v_\theta v_\phi)}
                {A_1B_3-A_3B_1},
\end{eqnarray}
which can now be integrated numerically.

\section{Boundary Condition for Low $\dot{M}$}

  Using the differential equations derived above, we can investigate the
boundary condition $v_{\theta,0}$ in more detail. This reinforces that
at low $\dot{M}$ the solutions naturally attract toward the boundary
condition of local viscous heating being balanced by local radiative
cooling. In the limit of $\theta\approx0$,
$g_{\rm eff}-g_{\rm rad}\approx g_{\rm eff}\approx GM/R^2-v_\phi^2/R$,
and $v_\phi\gg$ $\sqrt{kT/(\mu m_p)}$, $v_\theta$, the terms from above become
\begin{eqnarray}
        A_1 &=& v_\theta^2 - \frac{4}{5}\frac{kT}{\mu m_p}
		\approx - \frac{4}{5}\frac{kT}{\mu m_p},
        \\
        A_2 &=& 0,
        \\
        A_3 &=& \frac{4}{5}v_\theta
        \\
        B_1 &=& v_\theta^2
        \\
        B_2 &=& 0
        \\
        B_3 &=& \frac{14}{5}v_\theta
        \\
        C_1 &=& -F_\theta
        \\
        C_2 &=& -F_\phi
        \\
        C_3 &=& F_\phi - E_{\rm rad}.
\end{eqnarray}
In this limit, the derivative of $v_\theta$, equation (B10),  becomes,
\begin{eqnarray}
	\frac{dv_\theta}{d\theta}
	= \frac{-\frac{14}{5}F_\theta-\frac{4}{5}(F_\phi-E_{\rm rad})}
                {-\frac{52}{25}kT/(\mu m_p)-\frac{4}{5}v_\theta^2}
	\approx \frac{35F_\theta+10(F_\phi-E_{\rm rad})}{26kT/(\mu m_p)}.
\end{eqnarray}
Since $F_\theta \ll E_{\rm rad}$, $F_\phi$, the second term in the
numerator will dominate unless $E_{\rm rad}=F_\phi$. If $F_\phi>E_{\rm rad}$,
then $dv_\theta/d\theta>0$ and the solutions will be driven toward higher
$v_\theta$. From continuity, equation (24), this means that $y$ must decrease.
The radiative cooling is $E_{\rm rad}\propto g_{\rm rad}v_\theta\propto y^{-2}$
so that it will increase, pushing the solutions toward local energy balance.
If $E_{\rm rad}>F_\phi$, then $dv_\theta/d\theta<0$ and once again
local energy balance is reached for similar reasons.
The only time $dv_\theta/d\theta$ will vary slowly is when $E_{\rm rad}=F_\phi$
and we see in this case $dv_\theta/d\theta\approx35F_\theta/[26kT/(\mu m_p)]$,
which is positive. This is consistent with initial slope of $v_\theta$
in Figures (2) and (6). More detailed analysis, including the effects of
$g_{\rm rad}$, is needed to correctly predict the values of the slopes seen.

%

\end{appendix}

\end{document}